\documentstyle[prb,twocolumn,aps]{revtex}
\begin{document}
\draft
\title{\bf Inhibited Recombination of Charged Magnetoexcitons}
\author{H. Okamura,$^{1,2,3}$ D. Heiman,$^{2,3}$ M.
Sundaram,$^4$\cite{address} and A.C. Gossard$^4$}

\address{$^1$Physics Department, Kobe University,
Kobe 657-8501, Japan.}

\address{$^2$Physics Department, Northeastern University,
Boston, Massachusetts 02115}

\address{$^3$Francis Bitter Magnet Laboratory, Massachusetts
Institute of Technology, Cambridge, Massachusetts 02139}

\address{$^4$Materials Department,
University of California, Santa Barbara, California 93106}

\date{\today}
\maketitle
\begin{abstract}
Time-resolved
photoluminescence measurements show that the
decay time for charged excitons
in a GaAs two-dimensional electron
gas increases by an order of magnitude at high magnetic fields.
Unlike neutral excitons, the charged exciton center-of-mass is
spatially confined in a ``magnetically-adjustable quantum dot''
by the cyclotron orbit and the quantum well.
The inhibited recombination is
explained by a reduced phase coherence volume of the
magnetically-confined charged excitons.
\end{abstract}
\vspace{.5cm}

Charged excitons or ``trions'' were first identified in optical
absorption
experiments on electron-doped CdTe quantum well (QW) structures through
their polarization properties in a magnetic field.\cite{kheng}
The negatively charged exciton ($X^-$) transition in a narrow QW was
manifest in the spectra as a peak lying several
meV below the uncharged exciton ($X^0$) peak.  Although both
$X^0$ and $X^-$ transitions may had been observed previously
in PL spectra of GaAs QWs, the high electron density precluded their
identification.\cite{heiman88}  In hindsight, it is not surprising
that the recently identified $X^-$ is often the most common exciton
found in a system with excess electrons, similar to the $X^0$
in a system without excess electrons.
An $X^0$ in the presence of excess electrons becomes polarized
by a nearby electron and binds the electron by a dipolar attraction.
The properties of $X^-$ transitions in GaAs QWs have been explored
in several recent experimental\cite{buhmann,finkel,shields,gekhtman,yoon}
and theoretical\cite{wojs,palacios,chapman,liberman,whittaker} studies.

An interesting facet of the charged exciton
that has yet to be explored is the {\it confinement} produced
by the cyclotron motion in a magnetic field.
The $X^-$ complex (two electrons plus one hole) is singly-charged
and a magnetic field confines
the $X^-$ center-of-mass motion to a cyclotron orbit, unlike the
neutral exciton ($X^0$) which is free to move in a magnetic field.
This will be referred to as the
{\it magnetically-confined charged exciton} (MCX).
A magnetic field applied perpendicular to
a two-dimensional (2D) QW effectively confines the exciton to a
quantum dot (QD) whose size is adjustable with magnetic field.
The 3D MCX volume, defined roughly by the QW width
and the area of the cyclotron orbit in the plane of QW, is
inversely proportional to the perpendicular magnetic
field, $V_{_{MCX}} \propto 1/B_\perp$.  At high magnetic fields
this volume is typically smaller than QDs currently available
via patterned nanostructures.

The purpose of the present study is to examine the
radiative recombination of excitons confined in these MCX QDs.
Exciton recombination times were determined by measuring the
photoluminescence (PL) decay times in low-density GaAs/AlGaAs
electron gases in magnetic fields to 18 T, at temperatures
0.5-7~K.  At low temperatures, the $X^-$ decay time was found to
increase
by an order of magnitude for increasing perpendicular magnetic field.
In contrast, the recombination is rapid for both the $X^-$
in fields applied in the plane of the QW and for the uncharged $X^0$.
In the latter two cases the exciton is not confined to a QD.
The linear dependence of exciton decay time with magnetic field is
explained by a model in which the transition strength for optical
recombination is inversely proportional to the MCX QD volume or
phase coherence volume.

Experiments were performed on a symmetrically modulation-doped
electron gas contained in wide parabolic
GaAs/Al$_{0.3}$Ga$_{0.7}$As QWs.\cite{gossard}
In these QWs the electrons are distributed uniformly over a
thick layer $\sim$250~nm wide, with electron densities of 5
and 7 $\times$ 10$^{15}$~cm$^{-3}$, and mobility of
1.2 and 1.9~$\times$ 10$^5$~cm$^2$/Vsec.  The photo-generated
holes were confined within a layer $\sim$~25~nm
wide at the center of the much wider electron layer.
Thus although the electrons are spread over $\sim$~250~nm,
the excitons are confined to a narrow 2D plane in the center of
the QW.  Samples were mounted on a fiber optic probe inserted
into a $^3$He cryostat, which was placed in the bore of an 18~T
Bitter or a 30~T hybrid magnet.\cite{heiman92}
Time-resolved PL measurements
employed standard time-correlated single-photon counting
electronics and a multichannel plate.  A pulsed diode laser
operating at 1.58~eV (200~ps pulse length at 17~MHz) and
a 0.85~m double spectrometer provided a system response
of 300~ps full-width at half-maximum.\cite{manfred}
Deconvolution of
the system response resulted in a time resolution of
$\sim$ 20-100~ps for PL decay times.

Figure 1(a) shows the PL spectra at 0.5~K
in magnetic fields applied
perpendicular to the QW layers, and the inset (b) plots the
PL peak positions up to $B$=30~T.  There are two prominent PL
peaks, both showing a quadratic spectral shift at low fields
and a nearly linear shift at high fields, which is typical
of exciton emission.  Excitonic character of these PL lines
was further supported by the presence of a clear onset of
absorption in the PL excitation spectrum, and also by
strong resonant Rayleigh scattering.\cite{thesis}
In each spectrum the PL
peak at higher energy is assigned to recombination emission
from the $X^0$ neutral exciton and the peak at lower energy
is assigned to the $X^-$ charged exciton.
Assignment of the lower energy peak to the $X^-$ rather than to
a trapped exciton is in agreement with many other optical studies of
electron-doped GaAs QWs.\cite{buhmann,finkel,shields,gekhtman,yoon}
The singlet (antiparallel electron spins) and
triplet (parallel electron spins) states are not
resolved in this sample, however, another sample having
smaller PL linewidths showed $X^-$ peaks similar to those found
in a previous study of triplet $X^-$.\cite{shields}
The $X^-$ and $X^0$ peaks here
have strong opposite circular polarizations at high fields,
consistent with their peak assignments.
The energy separation between the two peaks is the binding
energy of the second electron, $\sim$~1~meV at high fields.
With increasing temperature the spectral intensity shifts
from the $X^-$ peak to the $X^0$ peak due to thermal
ionization of the second electron.  These spectral features
for the two peaks are quite similar to previous reports
of $X^-$ transitions in
GaAs QWs.\cite{buhmann,finkel,shields,gekhtman,yoon}

Results of time-resolved measurements at $T$=0.5~K are
shown in Fig.~2.  The inset displays the PL intensity of
the $X^-$ on a log scale as a function of time.  At $B$=0
the $X^-$ decay is rapid and closely follows the system
response (dashed curve).  Deconvolution of the system
response from the PL decay curve at $B$=0 yields a decay time
of $t \sim$~100~ps, which is close to that observed for
high-mobility 2D electron gas.\cite{heiman92,manfred,thesis}
For fields applied perpendicular to the QW,
the decay becomes increasingly longer at higher fields.
At $B$=18~T it reaches $t$=1.2~ns,
an order of magnitude longer than at $B$=0.
(Note that the field-induced increase in the PL decay time
is not due to changes in non-radiative decay channels,
as indicated by the nearly constant total intensity
of the two PL peaks in magnetic fields.)
Figure 2 plots PL decay times at
$T$=0.5 K for fields up to B=18 T.
The solid circles and solid squares
represent $X^-$ data measured for two samples with electron
densities differing by a factor of 1.4.
These data are nearly identical.
They demonstrate that the MCX {\it decay time is linearly
proportional to the magnetic field}.
In contrast, for fields applied parallel to the
QW, the $X^-$ decay time is independent of magnetic field,
shown by the open circles.  Furthermore, the decay time for the
$X^0$ peak (not shown) does not show appreciable lifetime
increase, and $t \leq 150$~ps for all fields.\cite{okamura94}
Rapid decay of $X^0$ was also found in a similar but undoped
sample (crosses in Fig. 2).\cite{thesis}

Combining these observations,
it is apparent that the inhibited recombination found at high
fields and low temperatures is only observed for the $X^-$,
and only for perpendicular magnetic field.  The four
cases of $X^0$ and $X^-$ with magnetic fields perpendicular
and parallel to
the QW layer are illustrated in Fig.~3(a).  For the $X^0$, the
exciton is confined only to the 2D layer since it is free
to move in the QW plane independent of the magnetic field
direction.  On the other hand, for the $X^-$ the exciton
complex is effectively 1D for $B$ parallel since
it is free to move along the field direction, yet is
0D for magnetic fields perpendicular to the QW layer.
Thus the only configuration giving complete confinement
in three dimensions is the latter case of $X^-$ with
$B$ perpendicular, in agreement with the
observed lifetime increase.

Concerning the free exciton recombination in QWs,
Feldmann {\it et al}\cite{feldmann} first pointed out the
importance of phase coherence volume for the
center-of-mass motion of excitons.
The coherence volume has a spatial extent
which is usually much larger than the Bohr-orbit size.
As a result, recombination radiation from free excitons emanates
from a volume in which the radiating exciton is phase coherent.
This phase coherence volume contains many unit cells radiating
coherently, producing a macroscopic polarization.
Thus the transition dipole strength and resulting radiative
decay rate is a linear function of the coherence
volume.\cite{feldmann,hanamura,andreani,eccleston,citrin}
Feldmann {\it et al.} demonstrated this linear relationship
experimentally through the dependence of exciton lifetime on
the temperature-dependent homogeneous spectral
linewidth.\cite{feldmann}   Later, a similar relationship
was demonstrated between exciton radiative lifetime and QD
size for CuCl microcrystals.\cite{itoh}
Below, we model the present situation of MCX based on this
coherence volume concept for excitons.

The characteristic volume for the magnetic
confinement of $X^-$ is given by
\begin{equation}
V_{_{MCX}} = L \cdot \pi l_B^2 \propto 1/B_\perp,
\end{equation}
where $L$ is the well width (hole layer width in the present
case), and $l_B = (c\hbar /eB_\perp)^{1/2}$ is the cyclotron
radius.   At small fields, $V_{_{MCX}}$ is larger than the
intrinsic coherence volume $V_0$ at zero field, and the oscillator
strength is determined by $V_0$ rather than $V_{_{MCX}}$.
At large fields, $V_{_{MCX}}$ becomes smaller than $V_0$,
limiting the spatial coherence of $X^-$.
The radiative decay time $\tau_r$ is then given by
\begin{equation}
\tau_r \propto \left\{ \begin{array}{cccc}
                    V_0^{-1} & \sim      & \mbox{constant}    & (\mbox{at small}~B_\perp) \\
                    V_{_{MCX}}^{-1} & \propto  & B_\perp   & (\mbox{at large}~B_\perp).
                       \end{array}
               \right.
\end{equation}
Although this simple model neglects effects of quantum
confinement and magnetic field on the internal
electron-hole ({\it e-h})
wave function of the exciton,
the results in (2) are in remarkably good qualitative agreement with the
data in Fig. 2, where $t$ shows large, nearly linear increases
at large $B_\perp$ fields, but it shows only small
variation at lower fields.

In reality, however, the exciton internal wave function
is strongly affected by quantum confinement and
magnetic field.   Effects of quantum
confinement were studied theoretically
by Takagahara\cite{takagahara}, taking into account the
valence band mixing and {\it e-h} exchange interaction.
He showed that the exciton oscillator
strength in a QD is a nearly linear function of the dot volume
even in the intermediate confinement regime,
primarily due to changing number of unit cells
in QDs of different sizes.   It has been known that
the effective Bohr radius of an exciton decreases
in a strong magnetic field,\cite{yafet} known as
``magnetic shrinkage''.
This would result in a larger
{\it e-h} wave function overlap,
and acting alone would produce a {\it shorter} decay time.
Clearly, this prediction is contrary to
the present observation of
decay time {\it increase} in magnetic fields.
Indeed, the observed $X^0$ decay time in an {\it undoped}
QW shows only small changes at high fields (see Fig. 2).
This demonstrates {\it experimentally}
that even though a magnetic field modifies the internal
wave function substantially, it does not lead to a
large change in decay time observed
in the present work.   It is unclear why the magnetic
shrinkage does not cause a
large change in the $X^0$ decay time.
However, for the confined $X^-$
the coherence volume effects could dominate over
the internal wave function effects, similar to the
case of quantum confinement.\cite{takagahara}
The above considerations nevertheless explain
the qualitatively good agreement between the
observed behaviors of PL decay time and our
model in spite of neglecting field- and confinement-induced
effects on the internal wave function.

Inhibited exciton decay could also have a contribution arising from
field-dependent changes of the mixing in the valence band ({\it vb})
states.   In the Luttinger model of QW {\it vb} levels, most levels
contain several wavefunction components having different angular
momentum states.\cite{broido,ekenberg}
For a given level, only certain components contribute to allowed
optical transitions, the other components do not contribute.
Thus if the magnetic field reduces the ratio of allowed to
unallowed components of a hole level, the exciton will
become ``dark'' and have a longer decay time.\cite{bawendi}
Calculations indicate that these changes are small,
typically much smaller than a factor of two.\cite{favrot}
These changes in {\it vb} mixing do not appear to account for the
order of magnitude increase observed in the decay time.

Figure 4 plots the measured PL decay times for both $X^-$ and
$X^0$ as a function of temperature. At $B$=14~T and
$T$=0.5~K, there is a large difference between
the decay times for $X^-$ and $X^0$ excitons.
This difference rapidly diminishes for increasing temperature,
and for $T\simeq$~3~K the two
decay times are nearly identical.  Data for $B$=6~T
show similar behavior but the effects are smaller.
These behaviors can be explained by considering the thermal
ionization of $X^-$.
There are two competing channels for $X^-$ decay,
namely
\begin{eqnarray}
X^- & \rightarrow & h\nu_- + e
\label{direct} \\
X^- & \rightarrow & X^0 + e  \rightarrow h\nu_0 +e ,
\label{ionization}
\end{eqnarray}
where $h\nu_-$ and $h\nu_0$ are the corresponding photons energies.
These decay channels are illustrated in Fig.~3(b).
In the direct process (\ref{direct}), an electron recombines
with the hole but leaves the remaining electron with
excess energy.
In the ionization process (\ref{ionization}), $X^-$ first
loses the excess electron to the electron gas by thermal
excitation, followed by a rapid $X^0$ recombination.
At low temperatures, the direct process is not thermally
activated, and most $X^-$ recombine directly with a long
decay time.
At elevated temperatures, in contrast,
the indirect decay channel (\ref{ionization}) also becomes
available for $X^-$, resulting in a rapid decrease of
the $X^-$ population.

The present work points out that charged excitons in a QW
are confined by a perpendicular magnetic field to
QDs having a size on the order of the cyclotron radius.
These QDs are magnetically-adjustable and their
volume is inversely proportional to the magnetic field.
The large increase in radiative lifetime of
the magnetically-confined charged exciton
is attributed to a decrease in its localization volume.
Similar changes in radiative lifetimes are expected for
other QW systems in high magnetic fields, and for nanofabricated
QD structures.

We thank G. Favrot and D. Broido for discussions on the
Luttinger model.
This work was supported by the National Science Foundation
through DMR-95-10699, and the work at UCSB received support
from QUEST and the NSF Science and Technology Center.

\begin{figure}
\caption{(a) Photoluminescence
spectra at $T$=0.5~K from a GaAs quantum well for various
magnetic fields applied perpendicular to the well.
The 250~nm wide electron layer had a density of
7~$\times$ 10$^{15}$~cm$^{-3}$; holes were confined
to a 25~nm wide well in the center of the electron layer.
(b) Spectral energies of the $X^-$ and $X^0$ peaks versus
magnetic field. }
\end{figure}

\begin{figure}
\caption{Photoluminescence decay time of excitons
in GaAs quantum well at $T$=0.5~K versus magnetic field.
The solid points are for the $X^-$ charged exciton
in perpendicular ($B_\perp$) magnetic field,
for electron densities of 5 (squares) and 7 (circles)
$\times$10$^{15}$~cm$^{-3}$.
Open circles are for $X^-$ in parallel ($B_\parallel$)
field.  The crosses are for the $X^0$
with $B_\perp$ in an undoped sample.  The solid, straight line
demonstrated the nearly linear
dependence of decay time on $B_\perp$.
The inset shows PL decay
at 0, 10 and 18~T, and system
response (SR, dashed curve).}
\end{figure}

\begin{figure}
\caption{(a) Schematic description for the center-of-mass motion of
neutral $X^0$ and charged $X^-$ excitons confined in a 2D plane for
parallel and perpendicular magnetic fields.
(b) Diagram of the two radiative processes for $X^-$ recombination:
the direct process of $X^-$ emission which leaves an electron behind;
and the indirect process in which the excess electron is first
freed by ionization before the $X^0$ recombines.}
\end{figure}

\begin{figure}
\caption{Photoluminescence decay time of the $X^-$ charged exciton
versus temperature for perpendicular magnetic fields of $B$=6 and 14~T,
for a GaAs quantum well.}
\end{figure}

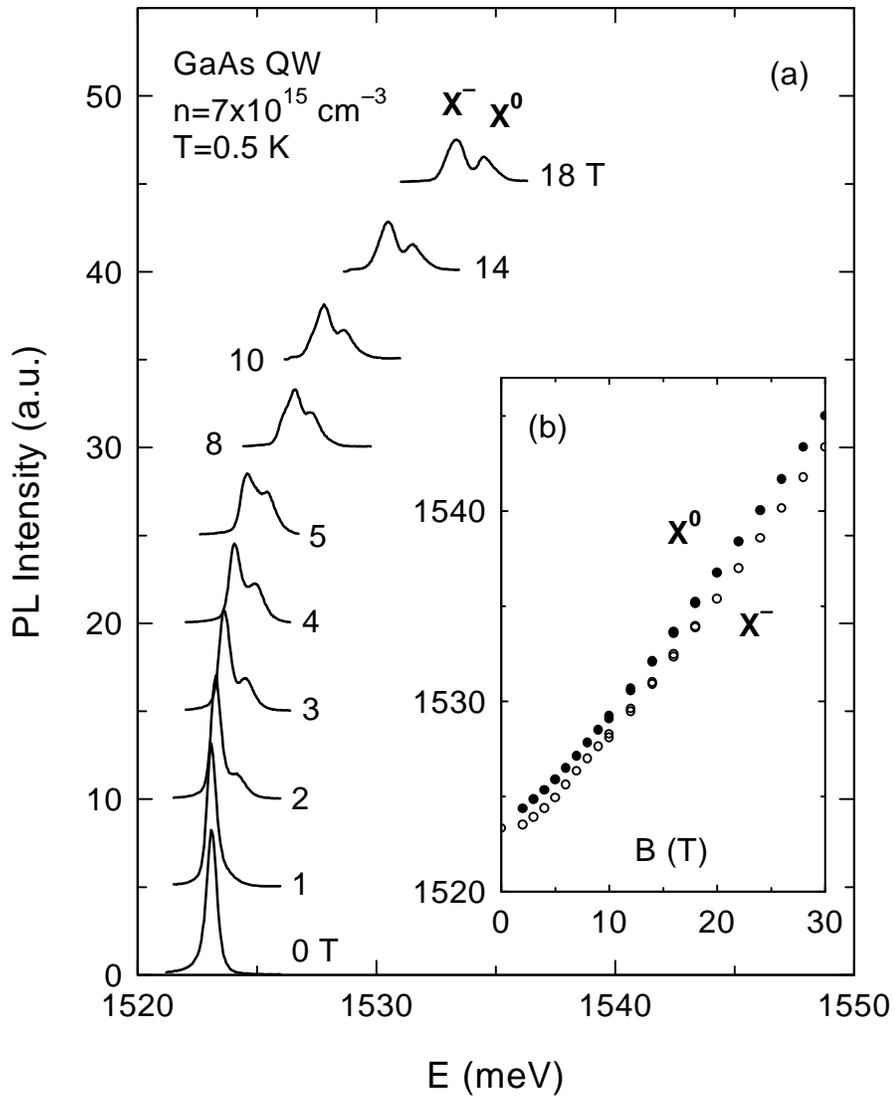
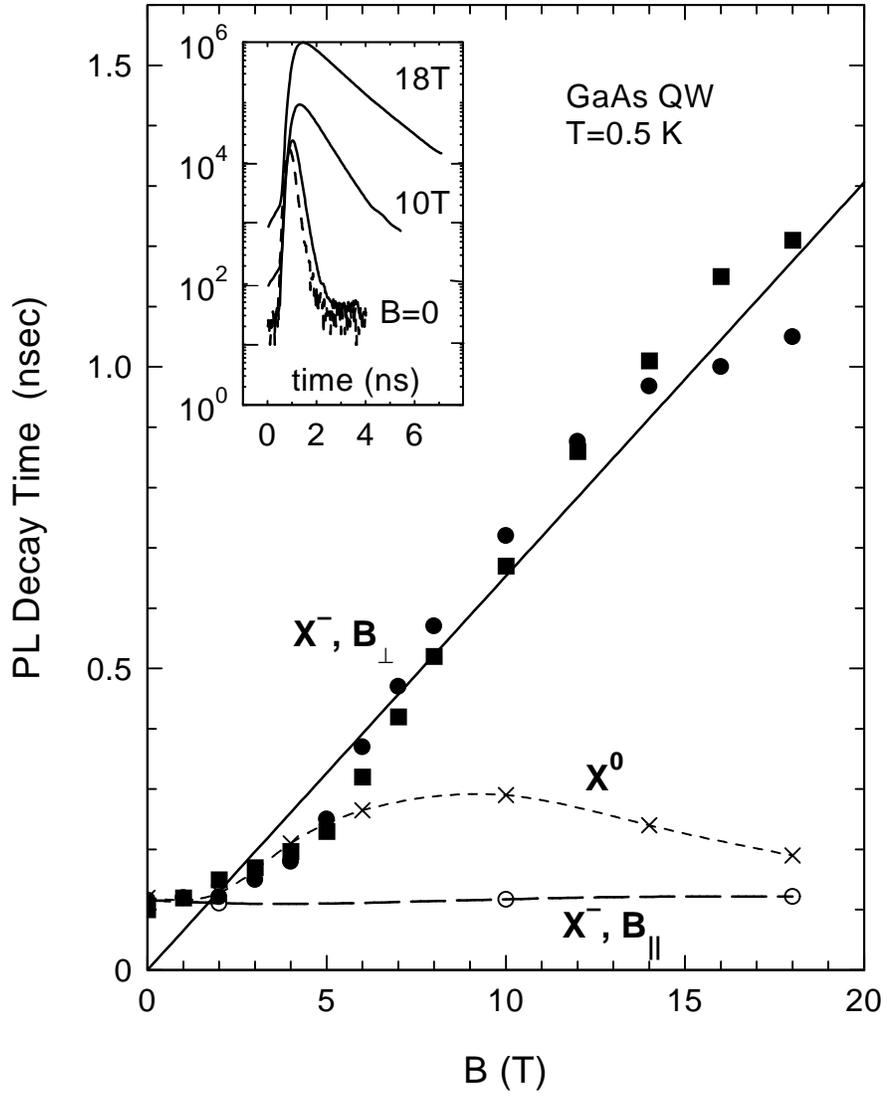
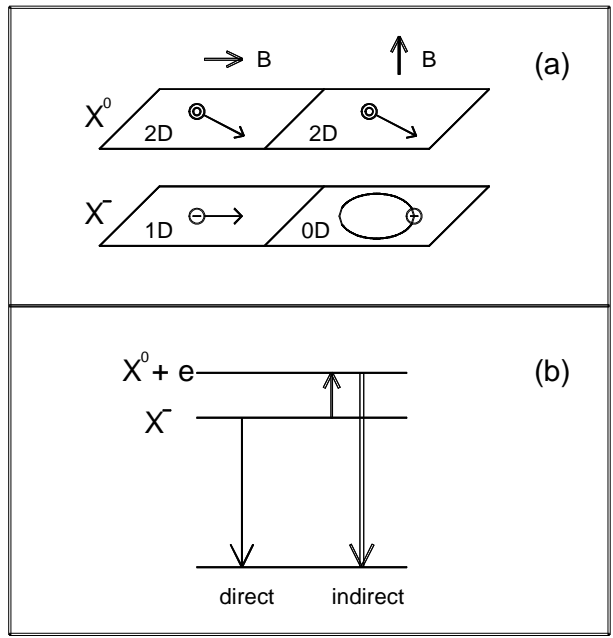
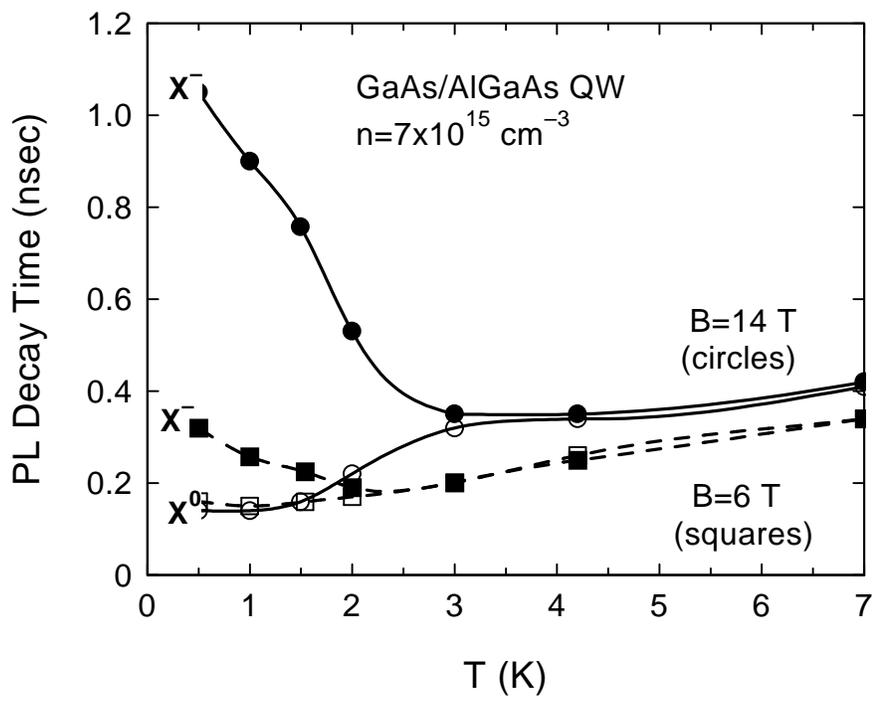
\end{document}